\documentclass[aip,apl,twocolumn,reprint,superscriptaddress,amsmath,amssymb]{revtex4-1}

\usepackage{graphicx, bm}
\usepackage{fp}
\usepackage[separate-uncertainty=true]{siunitx}
\usepackage{url}
\usepackage{hyperref}
\usepackage[sort&compress]{natbib}
\newcommand{\fig}[1]{Fig.~\ref{fig:#1}}
\newcommand{\eqn}[1]{Eq.~\ref{eqn:#1}}

\let\gammaraw\gamma
\renewcommand{\gamma}{\ensuremath{\frac{\gammaraw}{2\pi}}}
\renewcommand{\S}{\ensuremath{S_{11}}}
\newcommand{\Ms}{\ensuremath{M}}
\newcommand{\Meff}{\ensuremath{M_\mathrm{eff}}}
\newcommand{\geff}{\ensuremath{g_\mathrm{eff}}}
\newcommand{\kappac}{\ensuremath{\kappa_\mathrm{c}}}
\newcommand{\gammas}{\ensuremath{\gammaraw_\mathrm{s}}}
\newcommand{\omegac}{\ensuremath{\omega_\mathrm{c}}}
\newcommand{\omegaFMR}{\ensuremath{\omega_\mathrm{FMR}}}

\newcommand{\Tcomp}{\ensuremath{T_\mathrm{comp}}}
\newcommand{\muB}{\ensuremath{\mu_\mathrm{B}}}
\newcommand{\dS}{\ensuremath{\partial_B\S}}
\begin{document}

\title{Tunable magnon-photon coupling in a compensating ferrimagnet -- from weak to strong coupling}

\author{H. Maier-Flaig}
%\email{hannes.maier-flaig@wmi.badw.de}
\affiliation{Walther-Mei\ss ner-Institut, Bayerische Akademie der Wissenschaften, Garching, Germany}
\affiliation{Physik-Department, Technische Universit\"{a}t M\"{u}nchen, Garching, Germany}

\author{M. Harder}
\affiliation{Department of Physics and Astronomy, University of Manitoba, Winnipeg, Canada}

\author{S. Klingler}
\affiliation{Walther-Mei\ss ner-Institut, Bayerische Akademie der Wissenschaften, Garching, Germany}
\affiliation{Physik-Department, Technische Universit\"{a}t M\"{u}nchen, Garching, Germany}

\author{Z. Qiu}
\affiliation{WPI Advanced Institute for Materials Research, Tohoku University, Sendai, Japan}
\affiliation{Spin Quantum Rectification Project, ERATO, Japan Science and Technology Agency, Sendai, Japan}

\author{E. Saitoh}
\affiliation{WPI Advanced Institute for Materials Research, Tohoku University, Sendai, Japan}
\affiliation{Spin Quantum Rectification Project, ERATO, Japan Science and Technology Agency, Sendai, Japan}
\affiliation{Institute for Materials Research, Tohoku University, Sendai, Japan}
\affiliation{PRESTO, Japan Science and Technology Agency, Saitama, Japan}
\affiliation{Advanced Science Research Center, Japan Atomic Energy Agency, Tokai, Japan}

\author{M. Weiler}
\affiliation{Walther-Mei\ss ner-Institut, Bayerische Akademie der Wissenschaften, Garching, Germany}
\affiliation{Physik-Department, Technische Universit\"{a}t M\"{u}nchen, Garching, Germany}

\author{S. Gepr\"ags}
\affiliation{Walther-Mei\ss ner-Institut, Bayerische Akademie der Wissenschaften, Garching, Germany}
\affiliation{Physik-Department, Technische Universit\"{a}t M\"{u}nchen, Garching, Germany}

\author{R. Gross}
\affiliation{Walther-Mei\ss ner-Institut, Bayerische Akademie der Wissenschaften, Garching, Germany}
\affiliation{Physik-Department, Technische Universit\"{a}t M\"{u}nchen, Garching, Germany}
\affiliation{Nanosystems Initiative Munich, M\"{u}nchen, Germany}

\author{S. T. B. Goennenwein}
\affiliation{Walther-Mei\ss ner-Institut, Bayerische Akademie der Wissenschaften, Garching, Germany}
\affiliation{Physik-Department, Technische Universit\"{a}t M\"{u}nchen, Garching, Germany}
\affiliation{Nanosystems Initiative Munich, M\"{u}nchen, Germany}
\affiliation{Institut f\"ur Festk\"operphysik, Technische Universit\"at Dresden, Dresden, Germany.}

\author{H. Huebl}
\affiliation{Walther-Mei\ss ner-Institut, Bayerische Akademie der Wissenschaften, Garching, Germany}
\affiliation{Physik-Department, Technische Universit\"{a}t M\"{u}nchen, Garching, Germany}
\affiliation{Nanosystems Initiative Munich, M\"{u}nchen, Germany}

\date{\today}

\begin{abstract}
We experimentally study the magnon-photon coupling in a system consisting of the compensating ferrimagnet gadolinium iron garnet (GdIG) and a three-dimensional microwave cavity. The temperature is varied in order to tune the GdIG magnetization and to observe the transition from the weak coupling regime to the strong coupling regime. By measuring and modelling the complex reflection parameter of the system the effective coupling rate \geff\ and the magnetization \Meff\ of the sample are extracted. Comparing \geff\ with the magnon and the cavity decay rate we conclude that the strong coupling regime is easily accessible using GdIG. We show that the effective coupling strength follows the predicted square root dependence on the magnetization.
\end{abstract}


\maketitle
%\subsection{Introduction}
During the last decade, the coupling of paramagnetic moments with microwave photons has attracted a lot of attention.\cite{Xiang2013,Kurizki2015} 
%\subsection{Spin-Photon coupling}
Apart from the fundamental physical interest in the topic, coherent transfer of information between the two systems can be achieved when the coupling rate exceeds the individual relaxation rates.
This is of considerable interest for applications in quantum information processing as it opens the door to conversion between traveling photonic states and long-lived states in spin ensembles.\cite{Tkalcec2014} When the coupling rate between the sub-systems exceeds their invididual loss rates, the system enters the so-called strong coupling regime. 
While the coupling between an individual spin and an electromagnetic mode of a microwave resonator is typically in the weak coupling regime, the strong coupling regime can be reached by using a spin ensemble. The coupling strength is then boosted by a factor of $\sqrt{n}$ with $n$ being the number of polarized spins in the spin ensemble.\cite{Tavis1968,Fink2009,Imamoglu2009,Chiorescu2010}
Reaching the strong coupling regime using an ensemble of non-interacting (paramagnetic) spins and a microwave resonator is established in a variaty of configurations.\cite{Kubo2010,Abe2011,Schuster2010,Bushev2011} The $\sqrt{n}$ scaling of the coupling strength has been experimentally demonstrated by   \citet{Abe2011} and later \citet{Zollitsch2015}. In these experiments, the number of polarized spins $n$ was tuned by exploiting thermal depolarization of the spin ensemble. 
%\subsection{Magnon-Photon coupling}\noindent
Recently, the approach of strongly coupling a spin ensemble to a microwave resonator has been transferred to exchange coupled spin systems.\cite{Huebl2013,Zhang2014,Tabuchi2014,Bai2015,Kostylev2016} Since then, several experiments with increasing complexity have been implemented such as mediating the coupling of multiple independent ferromagnetic moments via a cavity \cite{Lambert2015,Zhang2015}, coupling a ferromagnetic material to a superconducting qubit \cite{Tabuchi2015} and probing strong coupling between spin system and a microwave resonator by optical means \cite{Klingler2016b,Hisatomi2016}. These experiments demonstrate the potential of ferromagnetic systems for magnon-microwave photon and magnon-optical photon conversion as well as memory applications even in the quantum limited regime. 
Interestingly, however, dedicated work on the foundation of the coupling mechanism, the $\sqrt{n}$ scaling of the coupling rate, does not exist for exchange coupled spin ensembles.
Here, we provide direct experimental proof of the $\sqrt{n}$ scaling in an exchange coupled ensemble of spins in a ferrimagnetic gadolinium iron garnet (GdIG) sample.

The theoretical description of the coupling of an ordered ferromagnet to an electromagnetic cavity has been discussed in various publications \cite{Soykal2010,Bai2015,Cao2015,Yao2015}. We use the theory developed by \citet{Cao2015} that approaches the problem in the 1D case starting from Maxwell's equations and describes the complex reflection parameter of a cavity loaded with a ferromagnet as:

\begin{equation}
\label{eqn:coupling}
\S = \frac{A\left(1 - \kappac\right)}
{i\left(\omega-\omegac \right ) - \kappac - i \geff^2\left(\omega-\omegaFMR+ i\gammas \right )^{-1}}.
\end{equation}
Here, \kappac\ and \gammas\ describe the decay rates (i.e. the half width at half maximum frequency line width) of the cavity and the magnon system, respectively. $A$ is a complex scaling parameter that accounts for losses and phase shifts in the setup. The effective coupling rate is denoted as \geff\ and the cavity resonance frequency is \omegac. The ferromagnetic resonance (FMR) frequency \omegaFMR\ depends sensitively on the magnetic anisotropy and the applied static magnetic field $H_0$. For thin ferromagnetic films with $H_0$ applied in the film plane, the dispersion is well described by the Kittel equation \cite{Kittel2005}
$\omega_\mathrm{FMR} = \gammaraw \mu_0 \sqrt{H_0\left(H_0+\Meff \right )}$.
Here, $\gammaraw$ is the gyromagnetic ratio of the material under investigation. The effective magnetization $\Meff = M- H_\mathrm{k}$ is equal to the magnetization \Ms\ if the shape anisotropy is the only relevant contribution to the anisotropy while $H_\mathrm{k}$ accounts for additional anisotropies such as magnetocrystalline anisotropy or strain induced anisotropy.
The effective coupling rate \geff\ is taken to be proportional to the square root of the net magnetic moment $m=M V$ of the sample. While in the 1D model of Ref.~\citenum{Cao2015} the magnetic moment is proportional to the thickness of the sample, in the 3D case considered here, it scales with the total volume $V$ of the sample. 
In contrast to paramagnets, the magnetization of ferromagnets typically shows only a weak temperature dependence for temperatures well below the Curie temperature. This weak temperature dependence is advantageous for applications as it makes the system more robust against external perturbations. In order to reach and study different coupling regimes, however, the sample size typically needs to be changed \cite{Tabuchi2014}. 

Here, we take a simple, robust and continuously tunable approach that allows for an in-situ manipulation of \geff\ by simply adjusting an external control parameter. We vary the net magnetic moment of a compensating \textit{ferri}magnet by an order of magnitude by changing temperature. 
Compensating ferrimagnets are a particular class of ferrimagnets containing two or more magnetic sublattices, where at least one of the  sublattices consists of internally weakly interacting moments, which can be thought of as acting like a paramagnetic spin ensemble. Therefore, thermal polarization and hence the net magnetization of this sublattice changes significantly with temperature. When this sublattice is antiferromagnetically coupled to the other sublattices, magnetization compensation can occur at the so-called compensation temperature \Tcomp.\cite{Dionne2009} At \Tcomp, the individual magnetizations of all involved sublattices cancel each other and the net remanent magnetization of the ferrimagnet vanishes. 
The macroscopic net magnetization that couples to the cavity photons can thus be tuned by temperature. 

Gadolinium iron garnet (GdIG) is a compensating ferrimagnetic insulator with three magnetic sublattices. Just like in the ubiquitous yttrium iron garnet, GdIG contains two iron sublattices that are strongly antiferromagnetically coupled and effectively form a single (net) iron sublattice with reduced magnetization. The magnetization of this iron sublattice shows only a weak temperature dependence below room temperature. A third sublattice is formed by the magnetic moments of the gadolinium ions that couple weakly to the iron moments and are aligned antiparallelly to the net iron magnetization. \cite{Dionne1979} The gadolinium sublattice magnetization shows a paramagnetic-like behavior, i.e. its magnetization increases towards low temperatures following a brillouin like function. At room temperature, the net magnetization of GdIG is dominated by the magnetization of the iron sublattices and therefore points along the larger Fe sublattice magnetization. However, as the magnetization of the Gd sublattice strongly increases with decreasing temperature, the net remanent magnetization decreases and vanishes at $\Tcomp$, where the sublattice magnetizations just compensate each other. Below \Tcomp\ the net magnetization increases again due to the increasing thermal polarization of the Gd sublattice and reaches a value of approximately \SI{595}{\kilo\ampere\per\meter} at \SI{5}{\kelvin}.\cite{Dionne2009} We investigate the latter temperature range in the following. 

%\subsection{Experiment, sample details}\noindent
\begin{figure}
\includegraphics[width=0.45\textwidth]{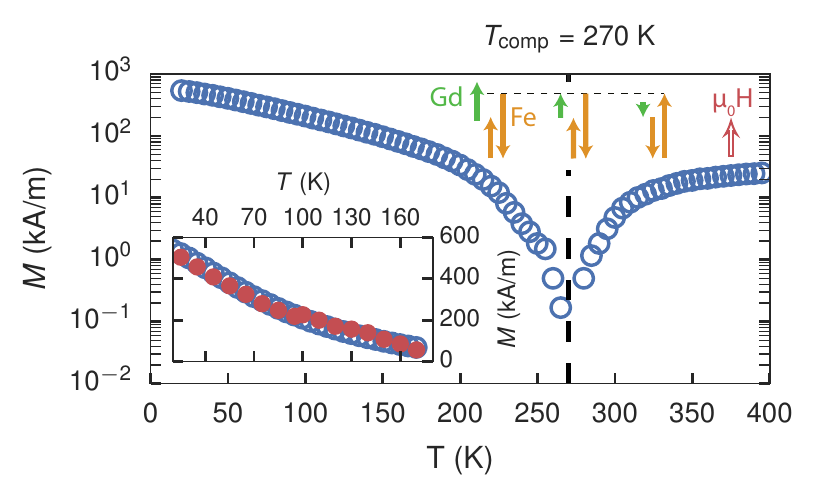}
\caption{Net magnetization \Ms\ measured at $\mu_0 H_0 = \SI{0.1}{\tesla}$ (applied in the film plane) using SQUID magnetometry (blue circles) normalized to the effective magnetization at \SI{15}{\kelvin}.
At the compensation temperature \Tcomp\ the net magnetization almost vanishes. Below \Tcomp, the Gd sublattice magnetization dominates the net magnetization and is therefore aligned with $H_0$.
Above \Tcomp, the net Fe magnetization dominates and aligns with $H_0$. The length of the arrows schematically represent the two Fe and the Gd sublattice magnetizations. Inset: Within the investigated temperature range, the evolution of the effective magnetization \Meff\ extracted from the FMR measurements (red data points) agrees very well with the net magnetization \Ms\ as determined by SQUID magnetometry.}
\label{fig:magnetization}
\end{figure}

The investigated sample is a $t=\SI{2.6}{\micro\meter}$ thick Gadolinium iron garnet film grown by liquid phase epitaxy on gadolinium gallium garnet (GGG) with lateral dimensions $l = \SI{5}{\milli\meter}$ and $b = \SI{2}{\milli\meter}$. The net magnetization \Ms\ of the sample was measured using SQUID magnetometry at an external magnetic field of \SI{0.1}{\tesla} (see \fig{magnetization}).\cite{Geprags2016} We find a compensation temperature of $\Tcomp = \SI{270}{\kelvin}$, which is lower than the bulk value of $\SI{285}{\kelvin}$\cite{Dionne2009} in agreement with literature suggesting that \Tcomp\ is slightly reduced in thin films.\cite{Pauthenet1958,Sawatzky1969}

For the magnon-photon coupling experiments, we place the sample in the magnetic field anti-node (electric field node) of the TE$_{011}$ mode of a 3D microwave cavity (Bruker Flexline MD5 dielectric ring cavity in an Oxford Instruments CF935 gas flow cryostat). 
The identical experimental setup was used in Ref.~\citenum{Maier-Flaig2016} which contains a more detailed description of the setup.

\begin{figure*}
\includegraphics[width=0.9\textwidth]{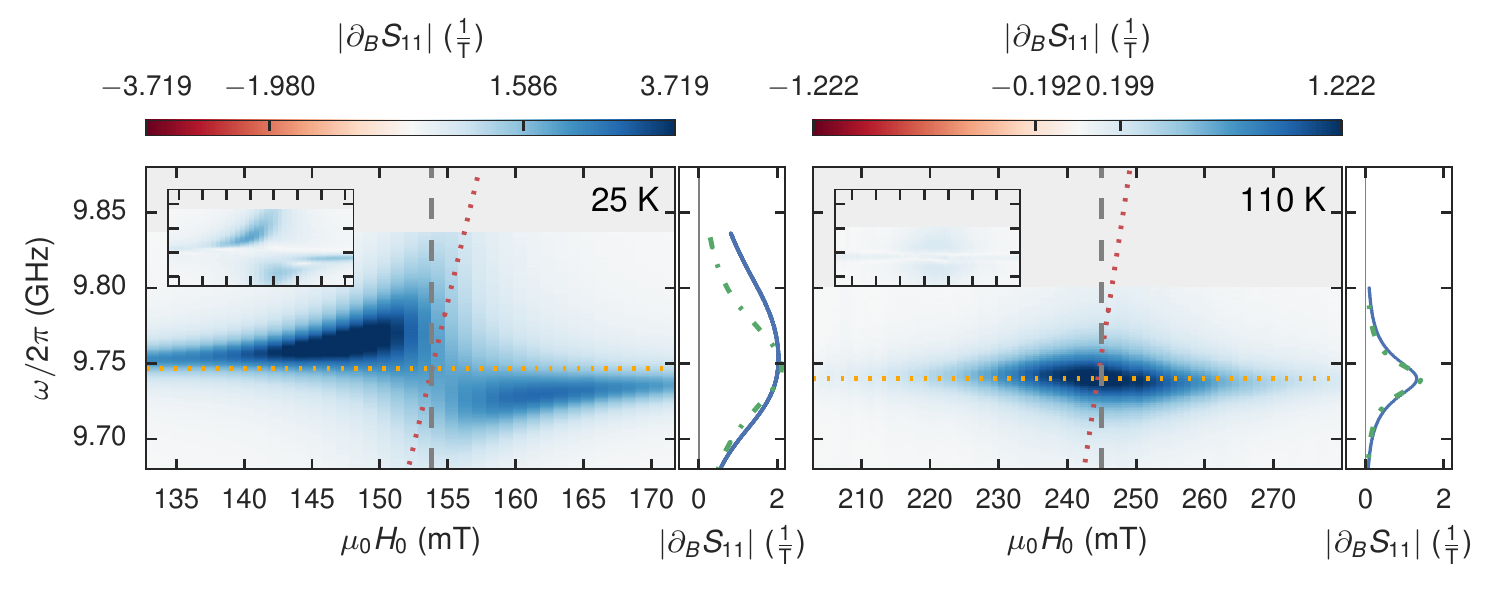}
\caption{Magnitude of the magnetic field derivative of the reflection parameter $\S$ at two distinct temperatures. The coupling visibly increases at low temperatures. 
The horizontal orange dotted line marks the resonance frequency $\omegac/2\pi$ of the unperturbed cavity, the red dotted line marks the resonance frequency of the unperturbed spin system $\omegaFMR/2\pi$.
Inset: Residuals of the fit to \eqn{coupling} on the same scale. 
%The ticks on the colorbar mark the maximum and minimum of the residual (smaller magnitude) resp. data (larger magnitude).
Line cuts: Data (blue) and fit (green line) at the field where the unperturbed cavity mode and magnon mode are degenerate (dashed vertical grey line in the adjacent colorplot).}
\label{fig:colorplot}
\end{figure*}

We measure the complex (phase sensitive) reflection scattering parameter \S\ around the resonance frequency of the cavity mode while applying a variable external magnetic field $H_0$ in the film plane. The applied probe power is chosen to be small ($0$~dBm) so that non-linear processes do not play a role. 
At this power, the number of photons excited in the cavity is approximately $N_\mathrm{Ph} = P/\left(\hbar \omegac \kappa_\mathrm{c}\right) = 2 \times 10^{13}$ and is several orders of magnitude lower than the minimum effective number of spins in the sample ($1 \times 10^{17}$ for the highest analyzed temperature).
In the microwave cabling connecting the cavity and the vector network analyzer, standing waves can form that cause a field independent background signal. In order to efficiently remove this background, we analyze the field derivative $\dS = \frac{\partial\S}{\partial (\mu_0 H_0)}$ as a function of $H_0$. Typical data for the absolute value of $\dS$ is shown in \fig{colorplot} for two distinct temperatures.
When the external magnetic field is adjusted to tune the FMR frequency $\omegaFMR$ close to the unperturbed cavity frequency $\omegac$, the excitations of the magnetic system (magnons) and the microwave cavity (photons) start hybridizing with a signature in \S\ described by \eqn{coupling}. For the case of strong coupling ($\geff\ > \kappac, \gammas$) [\fig{colorplot}~(\SI{25}{K})], \eqn{coupling} describes a characteristic anti-crossing of the $H_0$-independent cavity mode and the $H_0$-dependent spin resonance. The splitting can be used to determine the coupling strength.
In the weak coupling regime [\fig{colorplot}~(\SI{110}{K})], the cavity is only marginally disturbed. Here, the coupling strength can be determined by analyzing the change in the line width of the microwave cavity.\cite{Herskind2009} This case is equivalent to conventional FMR and therefore plotting the $S$ parameter at the cavity frequency as a function of the external field shows the typical FMR absorption line shape. Note that by taking the field derivative of \S, the unperturbed cavity absorption peak vanishes (\fig{colorplot}) as \omegac\ is field independent.

Because we aim to observe the transition of the system from the weak coupling regime ($\gammas > \geff$) to the strong coupling regime ($\kappac, \gammas \leq \geff$), the approximate solutions for the extreme cases are not sufficient. The system has to be modeled by the complete reflection characteristic given by \eqn{coupling}. To remove the field independent background of $S_{11}$, we first numerically calculate the magnetic field derivative of the complex \S\ parameter. We then perform a full 2D fit (i.e. all fits of all cuts at constant $H$ have shared parameter values) using the magnetic field derivative of \eqn{coupling} for every temperature. From this fit, we extract \Meff, \geff, \kappac\ and \gammas. The $g$-factor is fixed to $g=2$ over the whole temperature range based on Ref.~\citenum{Calhoun1958} thereby reducing the number of free parameters further. The resulting fit is virtually indistinguishable from the data in \fig{colorplot}. 
We therefore present the residual of the fit to the data in the insets on the same scale as the data. 
A vertical cut (dashed grey lines) of the 2D data and the fit is shown exemplarily for the static magnetic field corresponding to $\omegaFMR=\omegac$.
The 2D-fit is very good for high temperatures [\fig{colorplot}~(\SI{110}{K})], but the residuum shows some deviation of fit and data for low temperatures [\fig{colorplot}~(\SI{25}{K})]. We attribute this slight discrepancy to the presence of a second resonance line for our GdIG sample, which is apparent at low temperatures upon close examination. The second resonance might originate from spatial inhomogeneities in the sample. The data, the analysis scripts and results are publicly accessible under Ref.~\citenum{Maier-Flaig2017g} for further analysis and evaluation. 

We first discuss the effective magnetization \Meff\ extracted from the fitting procedure and displayed in the inset of \fig{magnetization} (red data points). The temperature evolution of the effective magnetization \Meff\ determined using FMR and the net magnetization \Ms\ measured by SQUID magnetometry agree well, indicating that the dominant anisotropy contribution in our GdIG thin film is indeed given by shape anisotropy. We therefore take $\Ms = \Meff$ in the following.  The observed slight difference of \Ms\ and \Meff\ can be explained by a small increase of the $g$-factor with decreasing temperature  as indicated in Ref.~\citenum{Calhoun1958}.

\begin{figure}
\includegraphics[width=0.45\textwidth]{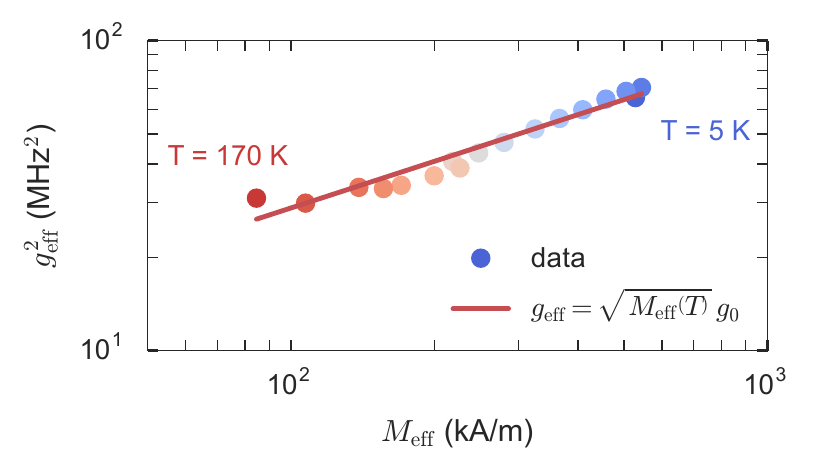}
\caption{Effective coupling rate $\geff^2$ as a function of the effective magnetization. The data agrees very well with the expected scaling behavior $\geff \propto \sqrt{\Meff}$ (red line).}
\label{fig:coupling}
\end{figure}

As central result, we confirm the scaling of the effective coupling rate \geff\ with the magnetic moment (or magnetization). As mentioned above, we expect $\geff \propto \sqrt{\Meff}$ in analogy to the paramagnetic case. \fig{coupling} shows $\geff^2$ as a function of \Meff\, where a straight line indicates the expected scaling  with $\geff=0$ at $\Meff=0$. The data follows this behavior accurately over an order of magnitude of the magnetization. As noted above, for low temperatures (large magnetization) a second resonance is indicated that we do not fit separately. The fit therefore slightly overestimates the coupling in this regime which can also be seen in \fig{coupling}. From the slope, we can calculate the single spin -- single photon coupling rate to $g_\mathrm{s}/2\pi = g_0/2\pi \left( \muB \sqrt{3}/V \right) ^{1/2} = \SI{0.072}{\hertz}$ with the Bohr magneton \muB\ and the sample volume $V$ and assuming a spin 1/2 particle with a $g$-factor of 2. This value is in reasonable agreement with the values of $g_\mathrm{s}/2\pi = \SI{0.043}{\hertz}$ determined for paramagnetic ensembles \cite{Abe2011}.

\begin{figure}
\includegraphics[width=0.45\textwidth]{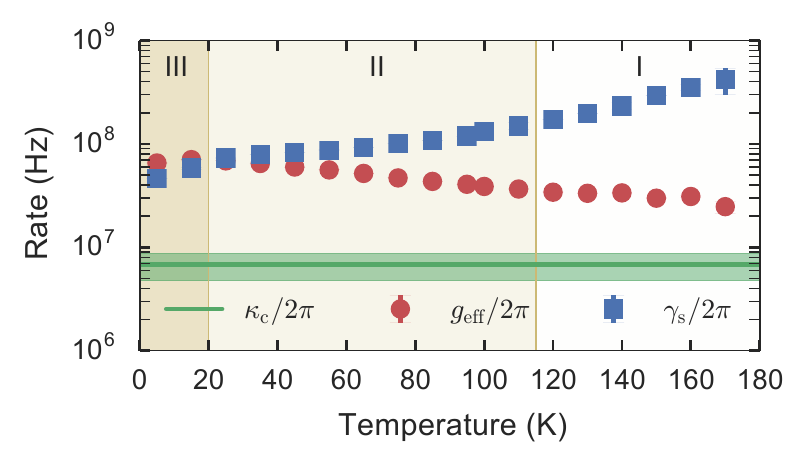}
\caption{Magnon relaxation rate $\gammas$, cavity relaxation rate $\kappac$ and effective coupling rate \geff\ as a function of temperature. The shaded green area denotes the standard error of the cavity relaxation rate. The coupling rate increases to low temperatures while the magnon relaxation rate drops. Thus, the system transitions from weak coupling (shaded region I) to a high cooperativity regime (region II) and enters strong coupling at low temperatures (region III).}
\label{fig:rates}
\end{figure}

In order to coherently exchange excitations between a magnonic system such as GdIG and a photonic system, strong coupling (i.e. $\kappac, \gammas \leq \geff$) is required. \fig{rates} displays the relaxation rate of the cavity \kappac\ and the spin system \gammas\ for comparison. The cavity decay rate \kappac\ should ideally be temperature independent but depends on the coupling of the feed line to the cavity.\cite{Maier-Flaig2016} As the temperature of the cavity is decreased, it thermally contracts slightly leading to a shift in resonance frequency and requires a change of the mechanical adjustment of the coupling mechanism. Hence, the decay rate of the cavity varies slightly with temperature. The mean value of \kappac\ is shown as green line in \fig{rates}, while its standard deviation is depicted as green shaded area. The FMR line width, \gammas\ significantly increases towards \Tcomp, in good agreement with reports in literature.\cite{Calhoun1958} Finally, \geff\ is plotted on the same scale and shows the increase towards low temperatures also shown in \fig{coupling}. Comparing the three rates, we find that the system is in the so-called high cooperativity regime ($\frac{\geff^2}{\kappac\gammas} > 1$) for temperatures below \SI{110}{\kelvin} (shaded region II in \fig{coupling}) and enters the strong coupling regime (region III) for liquid helium temperatures.
By choosing a slightly larger sample size or specially shaped cavity modes\cite{Kostylev2016}, strong coupling and thus the coherent exchange of information between a 3D cavity and GdIG is feasible even at higher temperatures.

%\subsection*{Conclusion}
In conclusion, we investigated the magnon-photon coupling in a system consisting of a compensating ferrimagnet (GdIG) and a 3D microwave cavity by measuring and analyzing the full complex reflection spectra quantitatively. We control the magnetization of GdIG with temperature and extract the scaling of the coupling strength with the net magnetization of the sample. We thereby confirm the expected scaling behavior $\geff = g_0 \sqrt{\Ms}$ of the ferrimagnet--cavity system. This result proves that the description used for paramagnets is equally appropriate for exchange coupled spin systems. The coupling strength for a system with vanishing net remanent magnetization, as found in compensating ferrimagnets at the compensation point or for the exchange resonances in anti-ferromagnets, poses an interesting non-linear problem\cite{Mukai2016} that has yet to be addressed.  In order to realize the transition from the strong to the weak coupling regime without modifying the sample or setup, we use a sample size that is just sufficient to reach the strong coupling regime. We emphasize, however, that strong coupling is easily accessible with GdIG / 3D microwave cavity system.

%\section*{Acknowledgements}
M. Harder acknowledges support from the NSERC MSFSS program. We gratefully acknowledge funding via the priority program Spin Caloric Transport (spinCAT), (Projects GO 944/4 and GR 1132/18), the priority program SPP 1601 (HU 1896/2-1) and the collaborative research center SFB 631 of the Deutsche Forschungsgemeinschaft.

\bibliography{references}

\end{document}